\documentclass[showpacs,amsmath,amssymb,aps,showkeys,floatfix,prd,a4paper]{revtex4}

\usepackage[dvips]{graphicx}
\usepackage{dcolumn}
\usepackage{bm}
\usepackage{epsfig}
\usepackage{amsfonts}
\usepackage{amssymb,amscd}

\def\lsim{\raise0.3ex\hbox{$<$\kern-0.75em\raise-1.1ex\hbox{$\sim$}}}

\def\gsim{\raise0.3ex\hbox{$>$\kern-0.75em\raise-1.1ex\hbox{$\sim$}}}

\newcommand{\be}{\begin{equation}}

\newcommand{\ee}{\end{equation}}

\def\beq{\begin{equation}}

\def\eeq{\end{equation}}

\def\beqa{\begin{eqnarray}}

\def\eeqa{\end{eqnarray}}

\newcommand{\ba}{\begin{eqnarray}}

\newcommand{\rr}{\mbox{\boldmath $r$}}

\newcommand{\rb}{\mbox{\boldmath $b$}}

\def\gappeq{\mathrel{\rlap {\raise.5ex\hbox{$>$}}

{\lower.5ex\hbox{$\sim$}}}}

\def\lappeq{\mathrel{\rlap{\raise.5ex\hbox{$<$}}

{\lower.5ex\hbox{$\sim$}}}}

\def\Toprel#1\over#2{\mathrel{\mathop{#2}\limits^{#1}}}

\newcommand{\rk}{\mbox{\boldmath $k$}}

\begin{document}

\title{Double vector meson production in the International Linear Collider}
\author{ F. Carvalho$^{1}$, V.P. Gon\c{c}alves $^{2}$,  B.D.  Moreira$^{3}$  and  F.S. Navarra$^3$}
\affiliation{$^1$ Departamento de Ci\^encias Exatas e da Terra, Universidade Federal de S\~ao Paulo,\\  
Campus Diadema, Rua Prof. Artur Riedel, 275, Jd. Eldorado, 09972-270, Diadema, SP, Brazil\\ 
$^{2}$ High and Medium Energy Group, Instituto de F\'{\i}sica e Matem\'atica,  Universidade Federal de Pelotas\\
Caixa Postal 354,  96010-900, Pelotas, RS, Brazil.\\
$^3$Instituto de F\'{\i}sica, Universidade de S\~{a}o Paulo,
C.P. 66318,  05315-970 S\~{a}o Paulo, SP, Brazil\\
}

\begin{abstract}
In this paper we study  double vector meson production in  $\gamma \gamma$ interactions at high energies and estimate, using the 
color dipole picture,  the main observables which can be probed  at the International Linear Collider (ILC). 
The total $\gamma (Q_1^2) + \gamma (Q_2^2) \rightarrow V_1 + V_2$ cross-sections  for $V_i = \rho$, $\phi$, $J/\psi$ 
and $\Upsilon$ are computed and the energy and virtuality dependencies are studied in detail. 
Our results demonstrate that the experimental analysis of this process  is feasible at the ILC 
and it can  be useful to constrain the QCD dynamics at high energies. 
\end{abstract}



\maketitle

\vspace{1cm}

\section{Introduction}

There is an increasing interest in the construction of a high energy electron-positron collider \cite{ilc}. The primary 
goal of this new facility will be to carry out precision measurements of electroweak physics, including the Higgs boson 
properties. An important byproduct of this program will be the study of high energy  photon-photon collisions \cite{serbo}
 and the continuation, at energies one order of magnitude higher, of the measurements performed at CERN-LEP, almost fifteen
 years ago. Photon-photon collisions are a  very clean  laboratory for the theory of strong interactions --  Quantum 
Chromodynamics (QCD) -- where we can test details of the QCD dynamics at high energies, such as  the evolution 
both in virtuality ($Q^2$) and in energy ($1/x$) (for a review see, e.g. Ref. \cite{nisius}). It has motivated the 
development of a large number of phenomenological studies in the last two decades 
\cite{Ginzburg,sectot_gg,Kwien_Motyka,nosfofo,serbo2,motyka,Qiao,dmvic1,dmvic2,Pire,vicmag07,Ivanov,
antoni}.  In particular, several authors have discussed the possibility of use the  scattering of two off-shell photons
 at high energy in $e^+\,e^-$ colliders  as a probe of the parton saturation effects in the QCD dynamics, which are 
predicted to be present in the high energy regime \cite{cgc}. Although the experimental results on  several  
inclusive and diffractive observables measured in $ep$ scattering at HERA and hadron - hadron collisions at RHIC 
and LHC suggest that these effects are already observed  in the energy regime probed by current colliders, these 
observations still need further confirmation. 

The state-of-art framework to treat  QCD at high energies is the Color Glass Condensate (CGC) formalism \cite{CGC2},
 which predicts the saturation of the growth of  parton distributions, with the evolution with the energy being 
described by an  infinite hierarchy of coupled equations for the correlators of  Wilson lines -- the Balitsky-JIMWLK
 hierarchy (for recent reviews see \cite{cgc}).  In the mean field approximation, this set of equations can be 
approximated by the Balitsky-Kovchegov (BK) equation \cite{BK}. As emphasized in Ref. \cite{nosfofo},  in general,  the 
applications of the CGC formalism to scattering problems require an asymmetric frame, in which the projectile has a simple
 structure and the evolution occurs in the target wave function, as it is the case in deep inelastic scattering. Therefore
 the extension of the  BK equation to the  calculation of the $\gamma \gamma$ scattering cross section is not a trivial task. 
In Ref. \cite{nosfofo} we have discussed this generalization in order to use  the solution of the BK equation as input of
 our calculations of the total $\gamma^* \gamma^*$ cross sections and photon structure functions, which were compared 
 with the LEP data. In particular, in Ref. \cite{nosfofo} we have improved  the treatment of the dipole - dipole cross
 section, which is the main ingredient of the description of the $\gamma \gamma$ interactions in the dipole picture. 
Differently from previous phenomenological studies, which disregarded the impact parameter dependence, we have 
proposed an educated guess for this dependence and demonstrated that the LEP data can be described in  this approach. 
The high energy behavior of the observables predicted in Ref. \cite{nosfofo} is largely different from those obtained 
in previous studies. This conclusion motivates us to 
review the analysis of other observables  which could be measured at the ILC.
One promising observable is  double vector meson production in $\gamma \gamma$ collisions, which has attracted the 
attention of several theoretical groups in the last years, with the cross section being estimated in different 
theoretical frameworks \cite{serbo2,motyka,Qiao,dmvic1,dmvic2,Pire,vicmag07,Ivanov,antoni}, as, for instance, the solution 
of the BFKL equation and impact factors at leading and next-to-leading orders.  
In this paper we will estimate the total $\gamma (Q_1^2) + \gamma (Q_2^2) \rightarrow V_1 + V_2$ cross-sections  for 
$V_i = \rho$, $\phi$, $J/\psi$ and $\Upsilon$ considering the improved treatment of the dipole - dipole cross section and  
the energy and virtuality dependencies of the total cross sections will be analyzed in detail. Our analysis  is strongly  
motivated by the fact that our knowledge about vector meson wave functions  has  improved considerably over the last 
years with the progress of phenomenological studies of vector meson production at HERA.  As a consequence the main ingredients of 
our calculations are constrained by  LEP and HERA data and hence our predictions for the ILC energies have only one free 
parameter -- the slope parameter $B_{V_1V_2}$ -- which determines the $t$ - dependence of the cross sections. The 
magnitude of this parameter for  different combinations of vector mesons is still  an open issue that deserves more detailed studies. 

This paper is organized as follows. In the next Section we present a brief review of the formalism, discussing in more detail 
the vector meson wave functions and the dipole - dipole cross section, which are the main inputs of our calculations. 
In Section \ref{results} we present our predictions for the production of different combinations of vector mesons. 
In particular, the dependencies of the cross sections on the energy and photon virtualities are analyzed in detail. 
Finally, in Section \ref{conc} we present our summary.

\section{Double vector meson production}
\label{formalismo}

\subsection{The cross section}

\begin{figure}
\centerline{\psfig{figure=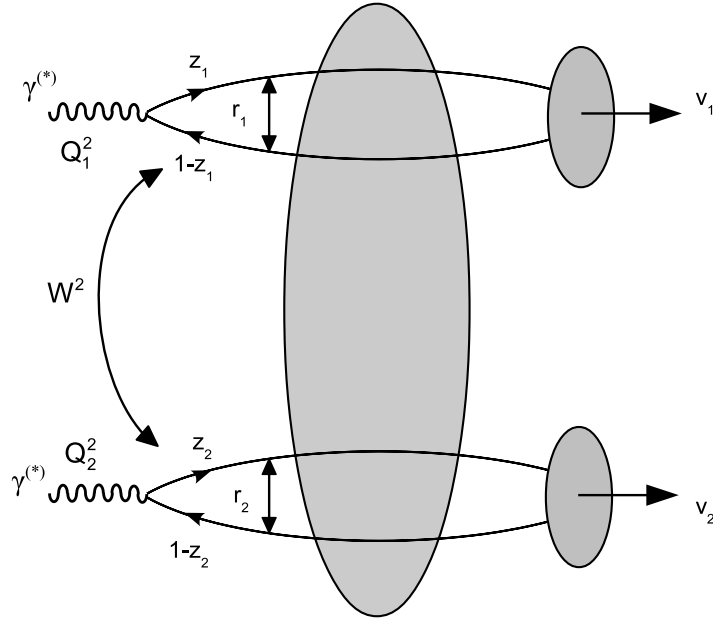,width=10cm}}
\caption{Double vector meson production in  $\gamma^* \gamma^*$ interactions at high energies in the 
color dipole picture.}
\label{fig1}
\end{figure}

Let us review the main formulas of vector meson
production in the color dipole picture (for more details see, e.g. Ref. \cite{vicmag07}). The relevant  
scattering process is  $\gamma^* \gamma^* \rightarrow V_1 \, V_2$, where $V_i$ stands for
both light and heavy vector mesons. At high energies, this scattering  can be seen 
as a succession in time of three factorizable subprocesses (See Fig. \ref{fig1}):  i) the photons fluctuate into  
quark-antiquark pairs (the dipoles), ii) these color dipoles interact and, iii) the pairs convert into the vector meson final states.
Using as kinematic variables the $\gamma^* \gamma^*$ c.m.s. energy
squared $s=W^2=(p+q)^2$, where $p$ and $q$ are the 
photon momenta,  the photon virtualities squared   given by 
$Q_1^2=-q^2$ and $Q_2^2 = -p^2$, and  $t$, the squared momentum transfer,
the total cross section for  double vector meson production is given by
\begin{eqnarray}
\sigma\, (\gamma \gamma \rightarrow V_1 \, V_2) = \int dt  \,  \frac{d\sigma  
(\gamma \gamma \rightarrow V_1 \, V_2)}{dt}\, =  \frac{1}{B_{V_1 \,V_2}} \left. \frac{d\sigma  
(\gamma \gamma \rightarrow V_1 \, V_2)}{dt}\,\right|_{t_{min}=0} =  \frac{[{\cal I}m \, {\cal A}(s,\,t=0)]^2}{16\pi\,B_{V_1 \,V_2}} \;,
\label{totalcs}
\end{eqnarray}
where we have  approximated the $t$-dependence of the differential cross section by an exponential  with  $B_{V_1 \, V_2}$ being 
the slope parameter. The imaginary part of the amplitude at zero momentum transfer ${\cal A}(s,\,t=0)$ reads as
\begin{eqnarray}
{\cal I}m \, {\cal A}\, (\gamma^* \gamma^* \rightarrow V_1 \, V_2) & = & \sum_{h, \bar{h}} \sum_{n, \bar{n}} 
\int dz_1\, d^2\rr_1 \,\Psi^\gamma_{h, \bar{h}}(z,\,\rr_1,\,Q_1^2)\,\, \Psi^{V_1*}_{h, \bar{h}}(z_1,\,\rr_1) \nonumber \\
&\times & \int dz_2\, d^2\rr_2 \,\Psi^\gamma_{n, \bar{n}}(z_2,\,\rr_2,\,Q_2^2)\,\, \Psi^{V_2 *}_{n, \bar{n}}(z_2,\,\rr_2)
\,
\sigma_{d d}(x_{12},\rr_1, \rr_2)
 \, ,
\label{sigmatot}
\end{eqnarray}
where $\Psi^{\gamma}$ and $\Psi^{V_i}$  are the light-cone wave functions  of the photon and vector meson, respectively. The quark and antiquark 
helicities are labelled by $h$, $\bar{h}$, $n$ and  $\bar{n}$ 
and reference to the meson and photon helicities are implicitly understood. The variable $\rr_1$ defines the relative transverse
separation of the pair (dipole) and $z_1$ $(1-z_1)$ is the longitudinal momentum fraction of the quark (antiquark). Similar definitions are valid 
for $\rr_2$ and  $z_2$. The variable $x_{12}$ will be defined  later. 
The basic blocks are the photon wave function, $\Psi^{\gamma}$, the  meson wave function, $\Psi_{T,\,L}^{V}$,  and the dipole-dipole  cross
section, $\sigma_{d\,d}$.

\subsection{Wave functions}

In the dipole formalism, the light-cone  wave functions $\Psi_{h,\bar{h}}(z,\,\rr)$ in the mixed 
representation $(\rr,z)$ are obtained through two dimensional Fourier 
transform of the momentum space light-cone wave functions $\Psi_{h,\bar{h}}(z,\,\rk)$ \cite{predazzi}.  This subject has been intensely discussed  
in several references  (see e.g.  Refs. \cite{Dosch:1996ss,stasto,sandapen,Kowalski:2003hm}). In what follows we present, for completeness, some of 
the main formulas. The  normalized  light-cone wave functions for longitudinally ($L$) and
 transversely ($T$) polarized photons are given by 
\begin{eqnarray}
\Psi^{L}_{h,\bar{h}}(z,\,\rr)& = & \sqrt{\frac{N_{c}}{4\pi}}\,\delta_{h,-\bar{h}}\,e\,e_{f}\,2 z(1-z)\, Q \frac{K_{0}(\varepsilon r)}{2\pi}\,,
\label{wfL}\\
\Psi^{T(\gamma=\pm)}_{h,\bar{h}}(z,\,\rr) & = & \pm
\sqrt{\frac{N_{c}}{2\pi}} \,e\,e_{f}
 \left[i e^{ \pm i\theta_{r}} (z \delta_{h\pm,\bar{h}\mp} -
(1-z) \delta_{h\mp,\bar{h}\pm}) \partial_{r}
+  m_{f} \,\delta_{h\pm,\bar{h}\pm} \right]\frac{K_{0}(\varepsilon r)}{2\pi}\,,
\label{wfT}
\end{eqnarray}
where $\varepsilon^{2} = z(1-z)Q^{2} + m_{f}^{2}$. The quark mass
$m_f$ plays the  role of a regulator when the photoproduction
regime is reached. The electric charge of the quark of flavor $f$ is given by $e\,e_f$.

One simple way to model the vector meson wave function is to assume, following Refs.~\cite{sandapen,Dosch:1996ss,Kowalski:2003hm}, 
that the vector meson is  a quark--antiquark state and that the spin and polarization structure is the same as in the photon case.  
The transversely polarized vector meson wave function is then given by
\begin{equation}
  \Psi^V_{h\bar{h},\lambda=\pm 1}(r,z) =
  \pm\sqrt{2N_c}\, \frac{1}{z(1-z)} \, 
  \left\{
  \mathrm{i}e^{\pm \mathrm{i}\theta_r}[
    z\delta_{h,\pm}\delta_{\bar h,\mp} - 
    (1-z)\delta_{h,\mp}\delta_{\bar h,\pm}] \partial_r \, + \, 
  m_f \delta_{h,\pm}\delta_{\bar h,\pm}
  \right\}\, \phi_T(r,z).
  \label{tspinvm}
\end{equation}
and the longitudinally polarized wave function is given by
\begin{equation}
  \Psi^V_{h\bar{h},\lambda=0}(r,z) = \sqrt{N_c}\,
  \delta_{h,-\bar h} \,
  \left[ M_V\,+ \, \delta \, \frac{m_f^2 - \nabla_r^2}{M_Vz(1-z)}\,  
    \right] \, \phi_L(r,z),
  \label{lspinvm}
\end{equation}
where $\nabla_r^2 \equiv (1/r)\partial_r + \partial_r^2$ and $M_V$ is the meson mass. 
The overlaps between the photon and the vector meson wave functions read then
\begin{align}
  (\Psi_V^*\Psi)_{T} &= \hat{e}_f e\, \frac{N_c}{\pi z(1-z)} \,
  \left\{m_f^2 K_0(\epsilon r)\phi_T(r,z) - \left[z^2+(1-z)^2\right]\epsilon K_1(\epsilon r) \partial_r \phi_T(r,z)\right\},
  \label{eq:overt}
  \\
  (\Psi_V^*\Psi)_{L} &=  \, \hat{e}_f e \, \frac{N_c}{\pi}\,
  2Qz(1-z)\,K_0(\epsilon r)\,
  \left[M_V\phi_L(r,z)+ \delta\,\frac{m_f^2 - \nabla_r^2}{M_Vz(1-z)}
    \phi_L(r,z)\right],
  \label{eq:overl}
\end{align}
where the effective charge $\hat{e}_f=1/3$, $2/3$,  $1/3$, or $1/\sqrt{2}$, for $\Upsilon$, $J/\psi$, $\Phi$, or $\rho$ mesons respectively.  
The  assumption that the quantum numbers of the meson are saturated by the quark--antiquark pair and that the possible contributions 
of gluon or sea-quark states to the wave function may be neglected, allows the normalization of the vector meson wave functions to unity.
%
%
The normalization conditions for the scalar parts of the wave functions are then 
\begin{gather}
  \label{eq:nnz_normt}
  1 = \frac{N_c}{2\pi}\int_0^1\!\frac{d{z}}{z^2(1-z)^2}\int\! d^2\vec{r}\;
  \left\{m_f^2\phi_T^2+\left[z^2+(1-z)^2\right]
  \left(\partial_r\phi_T\right)^2\right\},\\
  \label{eq:nnz_norml}
  1 = \frac{N_c}{2\pi} \int_0^1\!
  d{z}\,
  \int\! d^2\vec{r}\;
  \left[
    M_V\phi_L+
    \delta\,
    \frac{m_f^2-\nabla_r^2}{M_V z(1-z)}\,\phi_L\right]^2.
\end{gather}
Another important constraint on the vector meson wave functions is obtained from the decay width.  It is commonly assumed that the decay 
width can be described in a factorized way: the perturbative matrix element $q \bar{q} \to \gamma^* \to l^+ l^-$ factorizes out from 
the details of the wave function, which contributes only through its properties at the origin. 
The decay widths are then given by
\begin{gather}
  \label{eq:nnz_fvt}
  f_{V,T} = \hat{e}_f\, \left.\frac{N_c}{2\pi M_V}
  \int_0^1\!\frac{d{z}}{z^2(1-z)^2}
  \left\{m_f^2-\left[z^2+(1-z)^2\right]\nabla_r^2\right\}\phi_T(r,z)\right\rvert_{r=0},\\
  \label{eq:nnz_fvl}
  f_{V,L} = \hat{e}_f\, \left.\frac{N_c}{\pi}
  \int_0^1\!
  d{z}\,
  \left[
    M_V +\delta\, \frac{m_f^2-\nabla_r^2}{M_Vz(1-z)}\right]
  \phi_L(r,z)\right\rvert_{r=0}.
\end{gather} 
The coupling of the meson to the electromagnetic current, $f_V$, is obtained from the measured electronic decay width by
\begin{equation}
  \Gamma_{V\to e^+e^-} = \frac{4\pi\alpha_{\rm em}^2f_V^2}{3M_V}.
\end{equation}
We need now to specify the scalar parts of the wave functions, $\phi_{T,L}(r,z)$.   
Dosch, Gousset, Kulzinger and Pirner (DGKP) \cite{Dosch:1996ss} made the  assumption that the longitudinal momentum fraction $z$ 
fluctuates independently of the transverse quark momentum $\vec{k}$, where $\vec{k}$ is the Fourier conjugate variable to the dipole 
vector $\vec{r}$.  In the DGKP model one chooses $\delta=0$ in Eqs.  \eqref{eq:overl}, \eqref{eq:nnz_norml} and \eqref{eq:nnz_fvl}.  The DGKP model 
was further simplified by 
Kowalski and Teaney~\cite{Kowalski:2003hm}, who assumed that the $z$ dependence of the wave function for the longitudinally polarized 
meson is given by the short-distance limit of $z(1-z)$. For the transversely polarized meson they set 
$\phi_T(r,z) \propto [z(1-z)]^2$ in order to suppress the contribution from the end-points ($z\to 0,1$).  This leads to the ``Gauss-LC'' 
\cite{Kowalski:2003hm} wave functions given by: 
\begin{align}
  \phi_{T}(r,z) &= N_{T} [z(1-z)]^2\exp(-r^2/2R_{T}^2),\label{eq:Gaus-LC-T}\\
  \phi_{L}(r,z) &= N_{L} z(1-z)\exp(-r^2/2R_{L}^2).
  \label{eq:Gaus-LC-L}
\end{align}
The values of the constants $N_{T,L}$ and $R_{T,L}$ in Eqs. \eqref{eq:Gaus-LC-T} and \eqref{eq:Gaus-LC-L}, determined by requiring the correct 
normalization and by the condition $f_V=f_{V,T}=f_{V,L}$, are given in Table \ref{tab:GLCparams}. It is important to emphasize that this model allows 
to describe the HERA data and the recent LHC data for the exclusive vector meson photoproduction in hadron - hadron collisions 
(see, e.g. Refs. \cite{nos_prd,nos_plb,armeza}).
\begin{table}
  \centering
  \begin{tabular}{cccc|cccc}
    \hline\hline
    Meson & $M_V$/GeV & $f_V$ & $m_f$/GeV & $N_T$ & $R_T^2$/GeV$^{-2}$ & $N_L$ & $R_L^2$/GeV$^{-2}$ \\ \hline
    $\Upsilon$ &9.460& 0.236 & 4.5 & 0.67 & 2.16& 0.47 & 1.01 \\
    $J/\psi$ & 3.097 & 0.274 & 1.4 & 1.23 & 6.5 & 0.83 & 3.0  \\
    $\phi$ & 1.019 & 0.076 & 0.14 & 4.75 & 16.0 & 1.41 & 9.7  \\
    $\rho$ & 0.776 & 0.156 & 0.14 & 4.47 & 21.9 & 1.79 & 10.4 \\
    \hline\hline
  \end{tabular}
  \caption{Parameters of the ``Gauss-LC'' vector meson wave functions.}
  \label{tab:GLCparams}
\end{table}

\subsection{The dipole-dipole scattering cross section}

At  lowest order, the dipole - dipole interaction can be described by the two - gluon exchange between the dipoles, with the 
resulting cross section being energy independent (See, e.g. Ref. \cite{Navelet}). Taking into account the leading corrections 
associated to terms $\propto \log(1/x)$, as described by the BFKL equation, implies a power-law energy behaviour for the cross 
section, which violates the unitarity at high energies. These unitarity corrections were addressed in Ref. \cite{salam}, 
considering the color dipole picture and  independent multiple scatterings between the dipoles, and in Ref. \cite{iancu_mueller} 
considering the Color Glass Condensate formalism. In the eikonal approximation the dipole - dipole cross section can be expressed as follows
\begin{eqnarray}
\sigma^{dd} (\rr_1,\rr_2,Y)  = 2 \int d^2\rb \,{\cal{N}}(\rr_1,\rr_2,\rb,Y)
\end{eqnarray}
where  ${\cal{N}}(\rr_1,\rr_2,\rb,Y)$  is the scattering amplitude for the two dipoles with transverse sizes $\rr_1$ and $\rr_2$, 
relative impact parameter $\rb$ and rapidity separation $Y$. In Ref. \cite{nosfofo}    ${\cal{N}}$ was assumed to be given in terms 
of the solution of the BK equation obtained in Ref. \cite{bkrunning} disregarding the impact parameter dependence,  and we have proposed 
a model for the $\rb$ dependence, which limits the range of impact parameters which contribute to the cross section. The basic motivation
 for this model is associated to the fact that although the unitarity of the 
$S$-matrix (${\cal{N}} \le 1$) is respected by the solution of the BK equation (obtained disregarding the $\rb$ dependence), the 
associated dipole - dipole 
cross section can still rise indefinitely with the energy, even after the black disk limit (${\cal{N}} = 1$) has been reached at 
central impact parameters, due to the non-locality of the evolution. Consequently, a more elaborated model for the impact parameter 
dependence should be considered in order to obtain more realistic predictions for the dipole - dipole cross section. 
In Ref. \cite{nosfofo} we assumed that only the range $b < R$, where $R =$ Max$(r_1,r_2)$, contributes to the dipole - dipole cross section, 
{i.e.} we assumed that ${\cal{N}}$ is negligibly small when the dipoles have no overlap with each other ($b>R$). Therefore 
the dipole-dipole cross section can be expressed as follows \cite{nosfofo}: 
\begin{eqnarray}
\sigma^{dd} (\rr_1,\rr_2,Y)  = 2 \, {{N}}(\rr,Y) \int_0^R d^2\rb = 2 \pi R^2 {{N}}(\rr,Y)\,\,, 
\label{geral}
\end{eqnarray}
where ${{N}}(\rr,Y)$ is the solution of the BK equation obtained in Ref. \cite{bkrunning} disregarding the impact parameter dependence, 
which we denote rcBK hereafter.  The explicit form of $\sigma^{dd}$ reads
\begin{eqnarray}
\sigma^{dd} (\rr_1,\rr_2,Y) =  2 \pi r_1^2 N(r_2,Y_2) \, \Theta(r_1 - r_2) + 2 \pi r_2^2 N(r_1,Y_1) \, \Theta(r_2 - r_1) \,\,,
\label{ourmodel}
\end{eqnarray}
where  $Y_i = \ln (1/x_i)$ and 
\begin{eqnarray}
 x_i = \frac{Q_i^2 + 4 m_f^2}{W^2 + Q_i^2}.
\label{xdef}
\end{eqnarray}
As demonstrated in Ref. \cite{nosfofo}, using this model we can describe the LEP data for the total $\gamma \gamma$ cross sections and 
photon structure functions.

For comparison, in what follows we also will present the predictions obtained using  the phenomenological model for the dipole-dipole 
cross section proposed in \cite{Kwien_Motyka}. The inclusion of these predictions in our analysis, allows us to estimate the theoretical
 uncertainty present in ILC predictions, 
as well as to make comparisons  with   existing results in the literature. The dipole - dipole cross section proposed in Ref. \cite{Kwien_Motyka}
 is the following 
\begin{equation}
\sigma^{dd}_{a,b}(r_1,r_2,Y) = \sigma_0^{a,b}\,{{N}}(\rr_1,\rr_2,Y)
\label{sigmadd_mot}
\end{equation}
with $\sigma_0^{a,b} = (2/3) \sigma_0$, where $\sigma_0$ is a free parameter in the saturation model considered, fixed by fitting the DIS HERA 
data.  In the above equation  
${{N}}(\rr_1,\rr_2,Y) = {{N}}(\rr_{\rm\small  eff},Y=\ln(1/\bar x_{ab}))$, where 
\begin{equation}
r^2_{\rm\small  eff}\; = \;{r_1^2r_2^2\over r_1^2+r_2^2}\,\,\, \mbox{and}\,\,\, \bar x_{ab}
 \; = \;{Q_1^2 + Q_2^2 +4m_a^2+4m_b^2\over W^2+Q_1^2+Q_2^2}\,\,.
\label{reff}
\end{equation}
Moreover, as in Ref. \cite{nosfofo}, we also will present the predictions obtained using the phenomenological model for the 
forward dipole scattering $N(r,Y)$ proposed in Ref. \cite{iim}
and updated in \cite{soyez}, which was  constructed so as 
to reproduce two limits  of the LO BK equation 
analytically under control: the solution of the BFKL equation
for small dipole sizes, $r\ll 1/Q_s(x)$, and the Levin-Tuchin law 
for larger ones, $r\gg 1/Q_s(x)$. In the updated version of this parametrization \cite{soyez}, the free parameters were obtained by 
fitting the new H1 and ZEUS data.
In this parametrization the  forward dipole scattering amplitude is given by
\begin{eqnarray}
{{N}}(\rr,Y) =  \left\{ \begin{array}{ll} 
{\mathcal N}_0\, \left(\frac{r\, Q_s}{2}\right)^{2\left(\gamma_s + 
\frac{\ln (2/r Q_s)}{\kappa \,\lambda \,Y}\right)}\,, & \mbox{for $r 
Q_s(x) \le 2$}\,,\\
 1 - \exp^{-a\,\ln^2\,(b\,r\, Q_s)}\,,  & \mbox{for $r Q_s(x)  > 2$}\,, 
\end{array} \right.
\label{CGCfit}
\end{eqnarray}
where $a$ and $b$ are determined by continuity conditions at $\rr Q_s(x)=2$,  $\gamma_s= 0.6194$, $\kappa= 9.9$, $\lambda=0.2545$, 
$Q_0^2 = 1.0$ GeV$^2$,
$x_{0}=0.2131\times 10^{-4}$ and ${\mathcal N}_0=0.7$. Hereafter, we shall call the model above  IIM-S.
The first line from Eq. (\ref{CGCfit}) describes the linear regime whereas the second one includes saturation effects. 
One of the main motivations to use this model in our analysis is that it allows to estimate the magnitude of the saturation effects, 
by the comparison between the   predictions of the full model with those obtained considering only the linear term.

\section{Results}
\label{results}

In what follows, as in Ref. \cite{nosfofo},  we will denote the predictions obtained using the dipole - dipole cross section given by 
Eq. (\ref{sigmadd_mot}) by model 1 and those using  Eq. (\ref{ourmodel}) as input by  model 2. The parameters of our 
calculations are the same used in Ref. \cite{nosfofo} 
and this implies that our model gives a good description of the LEP data. As the value of the slope $B_{V_1V_2}$ for the different 
combinations of vector mesons in the final state  is still poorly known, we will, in almost all cases,  present our predictions for the 
product $B_{V_1V_2} \, \sigma(\gamma^*\gamma^* \rightarrow V_1V_2)$, which can be estimated without free parameters, since all 
parameters are constrained 
by the LEP and HERA data.

\begin{figure}
\centerline{\psfig{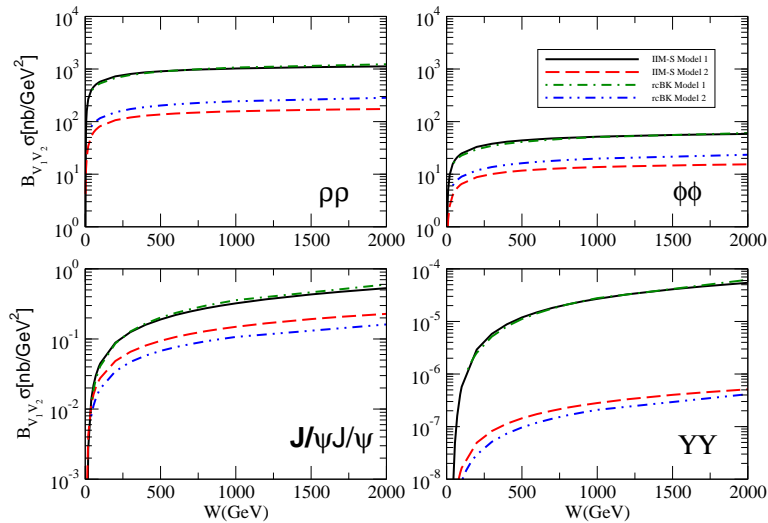}}
\caption{Energy dependence of the product $B_{V_1V_2} \, \sigma[\gamma^*(Q_1^2)\gamma^*(Q_2^2) \rightarrow V_1V_2]$ assuming 
$V_1 = V_2$ ($V_i = \rho, \phi, J/\Psi, \Upsilon$) and considering $Q_1^2 = Q_2^2 = 0$.}
\label{fig2}
\end{figure}

\begin{figure}
\centerline{\psfig{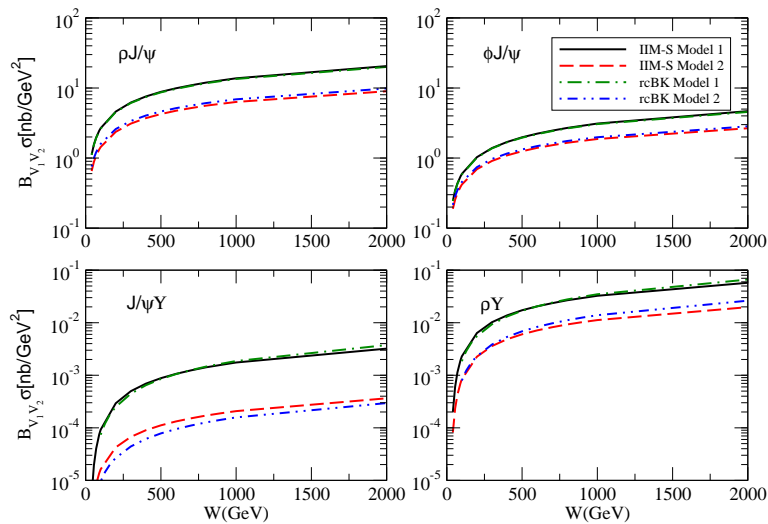}}
\caption{Energy dependence of the product $B_{V_1V_2} \, \sigma[\gamma^*(Q_1^2)\gamma^*(Q_2^2) \rightarrow V_1V_2]$ assuming 
$V_1 \neq V_2$ ($V_1 V_2 = \rho J/\Psi, \phi J/\Psi, J/\Psi \Upsilon, \rho \Upsilon$) and considering $Q_1^2 = Q_2^2 = 0$.}
\label{fig2b}
\end{figure}

In Fig. \ref{fig2} we present our predictions for the energy dependence of the product 
$B_{V_1V_2} \, \sigma(\gamma^*\gamma^* \rightarrow V_1V_2)$ assuming $V_1 = V_2$ ($V_i = \rho, \phi, J/\Psi, \Upsilon$) and
 considering $Q_1^2 = Q_2^2 = 0$. In this case only the transverse photon polarizations contribute to 
the total cross sections. It is important to emphasize that the color dipole picture allows us to treat simultaneously  double $\rho$ production by real 
photons, which is a typical soft process, and  double $\Upsilon$ production, which is the ideal laboratory to study the basic 
example of a hard process at high energies: the onium - onium scattering. Moreover, it allows us to study the transition between 
these two regimes, where we expect to see nonlinear 
(saturation) effects in the QCD dynamics. In our calculations we consider the two different models for the dipole - dipole cross 
section as well as the two models for the forward dipole scattering amplitude. We can observe that the main 
distinction is associated to the choice of the dipole - dipole cross sections. The predictions obtained using  model 2 are always 
smaller than 
those from  model 1. Previous estimates of the double vector production have overestimated the magnitude of the total cross sections. 
This behavior was expected from our previous results for the total $\gamma^*\gamma^*$ cross section \cite{nosfofo}.  
We also see that the difference between the predictions  increases with the quark masses, going from a factor 4, in the $\rho \rho$ case, 
to almost 
two orders of magnitude in the case of $\Upsilon \Upsilon$ production. This suggests that the experimental analysis of  double vector 
production at ILC can, 
in principle,  constrain the model for the dipole - dipole interaction. Moreover, in the case of  model 1, the IIM-S and rcBK predictions 
are almost 
identical for all combinations of vector mesons in the final state. In model 2  the IIM-S predictions are smaller than the rcBK ones for 
light vector meson production and larger  for  heavy vector meson production.  Such behavior is directly associated to the distinct transition 
between small and large dipoles predicted by these two models for the forward dipole scattering amplitude (see Fig. 2 in Ref. \cite{nosfofo}). 
As expected, we find that our predictions are strongly dependent on the quark mass, with the cross sections being smaller for the production 
of heavier vector mesons. Similar conclusions are obtained in the analysis  shown in  Fig. \ref{fig2b}, where we present our predictions for the 
energy dependence of the product $B_{V_1V_2} \, \sigma[\gamma^*(Q_1^2)\gamma^*(Q_2^2) \rightarrow V_1V_2]$ assuming 
$V_1 \neq V_2$ ($V_1 V_2 = \rho J/\Psi, \phi J/\Psi, J/\Psi \Upsilon, \rho \Upsilon$) and considering $Q_1^2 = Q_2^2 = 0$.

\begin{figure}
\centerline{\psfig{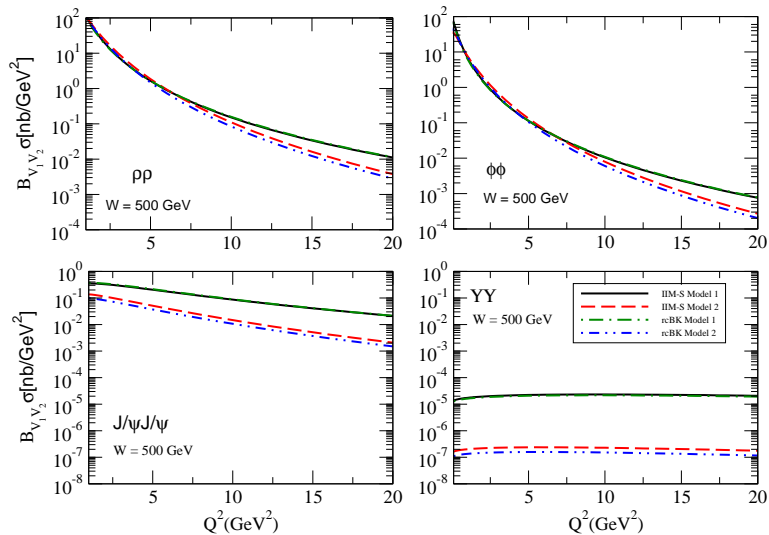}}
\caption{Dependence on the photon virtualities $Q_1^2 = Q_2^2 = Q^2$  of the product 
$B_{V_1V_2} \, \sigma[\gamma^*(Q_1^2)\gamma^*(Q_2^2) \rightarrow V_1V_2]$ assuming 
$V_1 = V_2$ ($V_i = \rho, \phi, J/\Psi, \Upsilon$) for a fixed center-of-mass energy ($W = 500$ GeV).}
\label{fig3}
\end{figure}

\begin{figure}
\centerline{\psfig{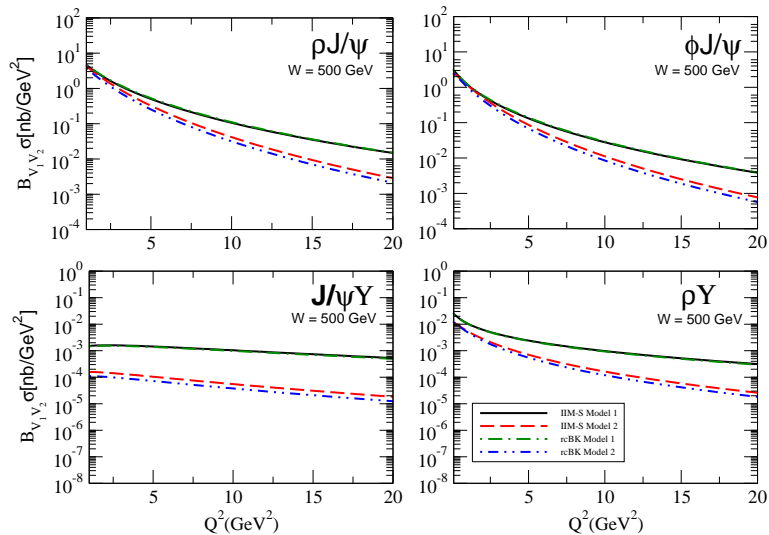}}
\caption{Dependence on the photon virtualities $Q_1^2 = Q_2^2 = Q^2$  of the product 
$B_{V_1V_2} \, \sigma[\gamma^*(Q_1^2)\gamma^*(Q_2^2) \rightarrow V_1V_2]$ assuming 
$V_1 \neq V_2$ ($V_1 V_2 = \rho J/\Psi, \phi J/\Psi, J/\Psi \Upsilon, \rho \Upsilon$) for a fixed center-of-mass  energy ($W = 500$ GeV). }
\label{fig3b}
\end{figure}

In  Fig. \ref{fig3}  we present our predictions for the dependence on the photon virtualities $Q_1^2 = Q_2^2 = Q^2$  of the product
 $B_{V_1V_2} \, \sigma[\gamma^*(Q_1^2)\gamma^*(Q_2^2) \rightarrow V_1V_2]$ for different combinations of vector mesons in the final
 state and  fixed center-of-mass energy ($W = 500$ GeV). In this case we  take into account the contributions of the transverse and
 longitudinal photon polarizations. For $Q^2 \neq 0$ we have two hard scales present in the process: the mass of the quarks 
(vector mesons) and the photon virtualities. For the double light vector meson ($\rho \rho, \phi \phi$) production, the dominant scale 
is the photon virtuality. 
In this case our predictions strongly decrease with $Q^2$. On the other hand, for the double $\Upsilon$ production, 
our predictions are almost $Q^2$ - independent in the range considered, since the dominant scale that defines the size 
of the two interacting  dipoles  is the bottom quark mass. In contrast, for the double $J/\Psi$ production, the characteristic 
dipole sizes are determined at small $Q^2$ by the charm quark mass and at medium $Q^2$ by the photon virtualities. Consequently, we observe 
a mild $Q^2$ dependence in the corresponding predictions. Moreover, we observe that the difference between  model 1 and model 2 predictions 
increases at larger $Q^2$ and for heavier vector mesons.  In Fig.  \ref{fig3b} we present our predictions for the production of two 
different vector mesons, which are similar to those observed in the production of identical vector mesons. Basically, the $Q^2$ 
dependence is reduced for larger values of the sum of the masses of the vector mesons  in the final state. 

\begin{figure}[t]
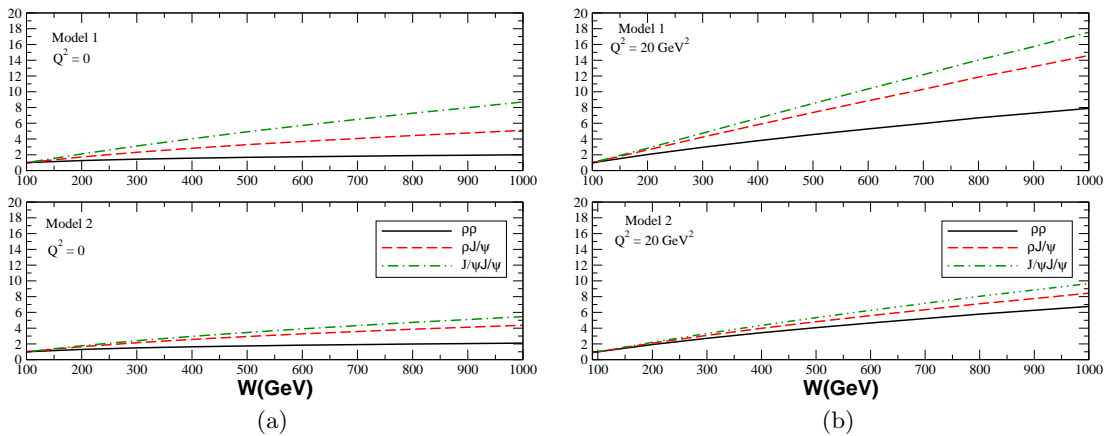

\begin{tabular}{ccc}
\psfig{figure=2meson_fig6a.eps,width=7cm}
& \,\,\, &
\psfig{figure=2meson_fig6b.eps,width=7cm}\\
(a) & \,\,\, &(b)
\end{tabular}
\caption{Energy dependence of the  normalized  cross sections (see text) for different
final states and different values of $Q_1^2 = Q_2^2 = Q^2$. (a) $ Q^2 = 0$ and  (b) $ Q^2 = 20$ GeV$^2$. }
\label{fig4}
\end{figure}

In order to illustrate how the energy behavior depends on the masses of the final state mesons, on the photon virtualities
 $Q_1^2 = Q_2^2 = Q^2$ and on the choice of the model for the dipole - dipole cross section, in Fig. \ref{fig4} we present 
our predictions for the normalized cross sections. The different cross sections were all normalized to the unity at 
$W = 100$ GeV  to better exhibit the different trends. For $Q^2 = 0$ we observe a clear transition between the soft and hard regimes, 
with the growth with the energy being faster for heavier mesons in the final state. 
Moreover, we find that  model 2 predicts a smaller slope than  model 1. For $Q^2 = 20$ GeV$^2$, a similar behavior is observed, 
but in this case already for the $\rho \rho$ production we see a steep rise of the cross section with the energy, which is directly 
associated to the presence  of the hard scale $Q^2$.

\begin{figure}
\centerline{\psfig{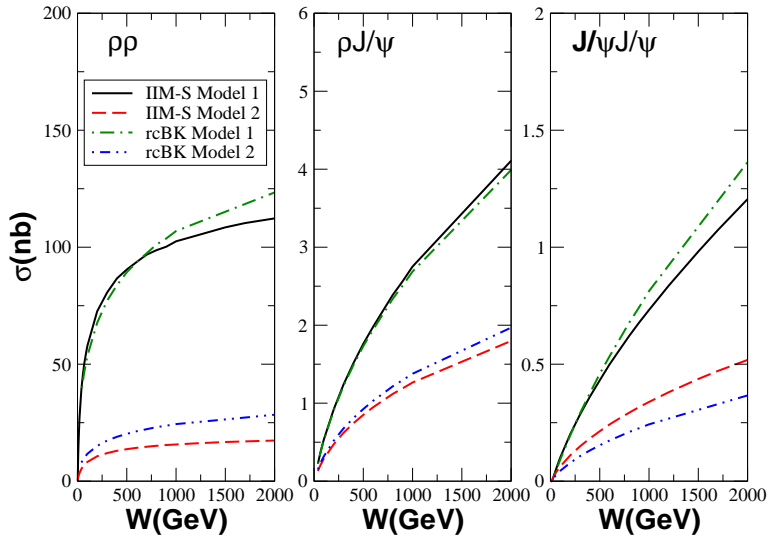}}
\caption{Energy dependence of the $\gamma \gamma \rightarrow V_1 V_2$ cross section for different final states considering $Q_1^2 = Q_2^2 = 0$.}
\label{fig5}
\end{figure}

In certain cases, where the slope parameters are phenomenologically known, it is possible to make definite predictions. 
In Fig. \ref{fig5}   we show the cross sections calculated with models 1 and 2 as a function of the energy $W$ with the proper slope 
coefficients, taken from \cite{vicmag07}: $B_{\rho \rho} = 10$ GeV$^{-2}$, $B_{\rho \psi} = 5$ GeV$^{-2}$ and $B_{\psi \psi} = 0.44$ GeV$^{-2}$. 
Our predictions with model 1 are similar to those obtained in  \cite{vicmag07}, with small differences mainly associated to the different 
forward dipole scattering amplitude and  to the treatment of the vector meson wave functions. In contrast, with model 2, we predict that at 
$W = 1$ TeV the cross sections are  $\sigma( \gamma \gamma \rightarrow \rho \rho) \approx 15$ nb,   
$\sigma( \gamma \gamma \rightarrow \rho J/\Psi) \approx 1.2$ nb and $\sigma( \gamma \gamma \rightarrow J/\Psi J/\Psi) \approx 0.25$ nb, 
which are a factor $\approx 4$ smaller than previous estimates in the literature obtained using the color dipole picture.

\begin{figure}
\centerline{\psfig{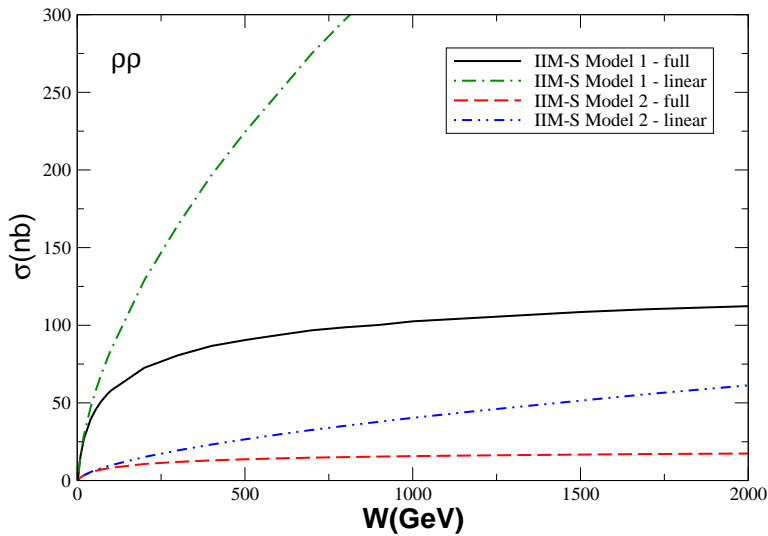}}
\caption{Comparison between the linear and full IIM-S predictions for the energy behavior of the
 $\gamma \gamma \rightarrow \rho \rho$ cross section. }
\label{fig6}
\end{figure}

Although  saturation effects of  QCD dynamics are expected to take over at higher energies, this change of dynamics 
can manifest itself with different  strength for different observables. For this reason, when working with the dipole approach,
 which is naturally prepared to incorporate  nonlinear corrections, it is always interesting to quantify the importance 
of the saturation effects. In Fig. \ref{fig6} we show our results for the energy dependence of the 
$ \gamma \gamma \rightarrow \rho \rho$ cross section, where we compare the full IIM-S model predictions with those obtained 
considering the linear regime of this model (first line in Eq. (\ref{CGCfit})). We see that the high energy behavior of 
the cross section is strongly modified by  saturation effects. This conclusion was already obtained in  \cite{vicmag07} and we see that it 
remains valid even after updating the dipole cross sections and model for  the dipole-dipole interaction.

\section{Summary}
\label{conc}

The scattering of two off-shell photons at high energy in $e^+\,e^-$ colliders is an interesting process to look for  parton saturation 
efffects. In these two-photon reactions, the photon virtualities can be made large enough to ensure the applicability of  perturbative 
methods or can be varied in order to test the transition between the soft and hard 
regimes of the QCD dynamics. In recent years, a series of studies have discussed in detail the treatment of the total cross section and 
the exclusive production of different final states in $\gamma \gamma$ interactions considering very distinct theoretical approaches. 
One great motivation for these works is the possibility that in a near future  $\gamma \gamma$ interactions may be investigated at 
the International Linear Collider (ILC). In particular, in Ref. \cite{nosfofo} we presented a detailed analysis of the $\gamma \gamma$
 cross section at high energies using the color dipole picture and taking into account  saturation effects, which are expected to be visible 
at high energies. In this paper we extended our approach to double vector meson production, improving  the  
previous analysis in three important aspects: i) the theoretical treatment of the dipole - dipole cross section; ii)  
the  forward scattering amplitude, considering  the solution of the running coupling BK equation (which is the state-of-art of the CGC 
formalism); and iii) the treatment of the vector meson wave functions. Considering that all parameters of our approach have been fixed by fitting  HERA and LEP data, our predictions 
for double vector meson production at ILC are  parameter free, except for  the only unknown parameter:   
the slope parameter $B_{V_1 V_2}$, which deserves a more detailed analysis.  
Our main conclusion is that the improvement of the theoretical framework for 
double  vector meson production in $\gamma \gamma$ interactions resulted in a reduction of the previously estimated cross sections at ILC energies. 
However, 
our results indicate that the experimental analysis at ILC is feasible and may be useful to constrain the QCD dynamics at high energies. 
As a final remark we would like to say that understanding vector meson production in $\gamma \gamma$ collisions is very important not 
only for the phenomenology of future electron-positron colliders but also for exclusive double vector meson production in hadron - hadron
 collisions, which have been studied at the LHC and that could be further studied in  future hadronic colliders.

\begin{acknowledgments}

This work was  partially financed by the Brazilian funding agencies CNPq, CAPES, FAPERGS and FAPESP.

\end{acknowledgments}

\hspace{1.0cm}


\begin{thebibliography}{99}



\bibitem{ilc}   
  G.~Moortgat-Pick, H.~Baer, M.~Battaglia, G.~Belanger, K.~Fujii, J.~Kalinowski, S.~Heinemeyer and Y.~Kiyo {\it et al.},
  arXiv:1504.01726 [hep-ph]; H. Baer, T. Barklow, K. Fujii, Y. Gao, A. Hoang, et al. arXiv:1306.6352 [hep-ph]; 
               M. Bicer et al., JHEP {\bf 1401}, 164 (2014) and references therein. 

\bibitem{serbo} 
  V.~M.~Budnev, I.~F.~Ginzburg, G.~V.~Meledin and V.~G.~Serbo,
  Phys.\ Rept.\  {\bf 15}, 181 (1975).

\bibitem{nisius}
  R.~Nisius,
  Phys.\ Rept.\  {\bf 332}, 165 (2000).



\bibitem{Ginzburg} 
  I.~F.~Ginzburg, S.~L.~Panfil and V.~G.~Serbo,
  Nucl.\ Phys.\ B {\bf 284}, 685 (1987).
  
  
\bibitem{sectot_gg}
J. Bartels, A. De Roeck, H. Lotter, { Phys. Lett. B } {\bf 389},
742 (1996);   A.~Donnachie, H.~G.~Dosch and M.~Rueter,
  Phys.\ Rev.\ D {\bf 59}, 074011 (1999); 
J. Bartels, C. Ewerz, R. Staritzbichler, { Phys. Lett.
B } {\bf 492}, 56 (2000);
 A. Bialas, W. Czyz, W. Florkowski, Eur. Phys. J. C
{\bf 2}, 683 (1998); J. Kwiecinski, L. Motyka, Eur. Phys. J. C
{\bf 18}, 343 (2000);   N. N. Nikolaev, B. G. Zakharov, V. R.
Zoller, JETP {\bf 93}, 957 (2001);
S. J. Brodsky, F. Hautmann, D. E. Soper, Phys. Rev. D {\bf 56},
6957 (1997); Phys. Rev. Lett. {\bf 78}, 803 (1997);
M. Boonekamp, A. De Roeck, C. Royon, S. Wallon, Nuc. Phys. B {\bf
555}, 540 (1999);  S.J. Brodsky, V.S. Fadin, V.T. Kim, L.N. Lipatov, 
G.B. Pivovarov, Pis'ma ZHETF {\bf 76}, 306 (2002) [JETP Letters {\bf 76}, 249 (2002)];   M.~Kozlov and E.~Levin,
  Eur.\ Phys.\ J.\ C {\bf 28}, 483 (2003); 
  V.~P.~Goncalves, M.~V.~T.~Machado and W.~K.~Sauter,
  J.\ Phys.\ G {\bf 34}, 1673 (2007); 
  F.~Caporale, D.~Y.~Ivanov and A.~Papa,
  Eur.\ Phys.\ J.\  C {\bf 58}, 1 (2008);  D.~Y.~Ivanov, B.~Murdaca and A.~Papa,
                JHEP {\bf 1410}, 58 (2014); G.~A.~Chirilli and Y.~V.~Kovchegov,
  JHEP {\bf 1405}, 099 (2014).
   
\bibitem{Kwien_Motyka}
N. Timneanu, J. Kwiecinski, L. Motyka,  Eur. Phys. J. C {\bf 23}, 513 (2002).


\bibitem{nosfofo} V.~P.~Goncalves, M.~S.~Kugeratski, E.~R.~Cazaroto, F.~Carvalho and F.~S.~Navarra,
                  Eur.\ Phys.\ J.\ C {\bf 71}, 1779 (2011).


\bibitem{serbo2} 
  I.~F.~Ginzburg, S.~L.~Panfil and V.~G.~Serbo,
  Nucl.\ Phys.\ B {\bf 296}, 569 (1988).

\bibitem{motyka} 
  J.~Kwiecinski and L.~Motyka,
  Phys.\ Lett.\ B {\bf 438}, 203 (1998).

\bibitem{Qiao} 
  C.~F.~Qiao,
  Phys.\ Rev.\ D {\bf 64}, 077503 (2001).  
  
\bibitem{dmvic1} 
  V.~P.~Goncalves and M.~V.~T.~Machado,
  Eur.\ Phys.\ J.\ C {\bf 28}, 71 (2003);
  Eur.\ Phys.\ J.\ C {\bf 29}, 271 (2003). 

\bibitem{dmvic2} 
  V.~P.~Goncalves and W.~K.~Sauter,
  Eur.\ Phys.\ J.\ C {\bf 44}, 515 (2005); 
  Phys.\ Rev.\ D {\bf 73}, 077502 (2006).

\bibitem{Pire} 
  B.~Pire, L.~Szymanowski and S.~Wallon,
  Eur.\ Phys.\ J.\ C {\bf 44}, 545 (2005); 
  R.~Enberg, B.~Pire, L.~Szymanowski and S.~Wallon,
  Eur.\ Phys.\ J.\ C {\bf 45}, 759 (2006)
  [Erratum-ibid.\ C {\bf 51}, 1015 (2007)]; 
  M.~Segond, L.~Szymanowski and S.~Wallon,
  Eur.\ Phys.\ J.\ C {\bf 52}, 93 (2007).


\bibitem{vicmag07} V.~P.~Goncalves and M.~V.~T.~Machado,
                     Eur.\ Phys.\ J.\ C {\bf 49}, 675 (2007).

\bibitem{Ivanov} 
  D.~Y.~Ivanov and A.~Papa,
  Eur.\ Phys.\ J.\ C {\bf 49}, 947 (2007);  Nucl.\ Phys.\ B {\bf 732}, 183 (2006). 

\bibitem{antoni} 
  M.~Klusek, W.~Schafer and A.~Szczurek,
  Phys.\ Lett.\ B {\bf 674}, 92 (2009);  S.~Baranov, A.~Cisek, M.~Klusek-Gawenda, W.~Schafer and A.~Szczurek,
                     Eur.\ Phys.\ J.\ C {\bf 73}, 2335 (2013).

      



\bibitem{cgc}  F.~Gelis, Int.\ J.\ Mod.\ Phys.\ A {\bf 28}, 1330001 (2013); 
               F.~Gelis, E.~Iancu, J.~Jalilian-Marian and R.~Venugopalan, arXiv:1002.0333;
               E.~Iancu and R.~Venugopalan, arXiv:hep-ph/0303204; 
               H.~Weigert,  Prog.\ Part.\ Nucl.\ Phys.\  {\bf 55}, 461 (2005); 
               J.~Jalilian-Marian and Y.~V.~Kovchegov, Prog.\ Part.\ Nucl.\ Phys.\  {\bf 56}, 104 (2006).



\bibitem{CGC2}
L.~D.~McLerran and R.~Venugopalan,
Phys.\ Rev.\ D {\bf 49}, 2233 (1994);
E. Iancu, A. Leonidov, L. McLerran, Nucl. Phys. A {\bf 692}, 583
(2001); E. Ferreiro, E. Iancu, A. Leonidov, L. McLerran, Nucl.
Phys. A {\bf 703}, 489 (2002);J. Jalilian-Marian, A. Kovner, L.
McLerran  and  H. Weigert, Phys. Rev. D {\bf 55}, 5414 (1997); J.
Jalilian-Marian, A. Kovner and  H. Weigert, Phys. Rev. D {\bf 59},
014014 (1999), {\it ibid.} {\bf 59}, 014015 (1999), {\it ibid.}
{\bf 59}  034007 (1999); A. Kovner, J. Guilherme Milhano and  H.
Weigert,   Phys. Rev. D {\bf 62},  114005 (2000);
H. Weigert, Nucl. Phys.  {\bf A703}, 823 (2002).


\bibitem{BK}
I.~Balitsky,
Nucl.\ Phys.\ B {\bf 463}, 99 (1996);
Y.~V.~Kovchegov,
Phys.\ Rev.\ D {\bf 60}, 034008 (1999);
Phys.\ Rev.\ D {\bf 61}, 074018 (2000).



\bibitem{predazzi}  V.~Barone and E.~Predazzi,
\textit{High-Energy Particle Diffraction}, Springer-Verlag, Berlin Heidelberg, (2002).

\bibitem{Dosch:1996ss}
  H.~G.~Dosch, T.~Gousset, G.~Kulzinger and H.~J.~Pirner,
  Phys.\ Rev.\ D {\bf 55},  2602  (1997).



\bibitem{stasto}
S.~Munier, A.~M.~Stasto and A.~H.~Mueller,
Nucl.\ Phys.\ B {\bf 603}, 427 (2001).


\bibitem{sandapen}
J.~R.~Forshaw, R.~Sandapen and G.~Shaw,
 Phys. Rev. D {\bf 69}, 094013 (2004).

\bibitem{Kowalski:2003hm}  H.~Kowalski and D.~Teaney,
                           Phys.\ Rev.\ D {\bf 68}, 114005 (2003);  H.~Kowalski, L.~Motyka and G.~Watt,
                           Phys.\ Rev.\  D {\bf 74},  074016 (2006).
       
       

\bibitem{nos_prd} V.~P.~Goncalves, B.~D.~Moreira and F.~S.~Navarra, Phys.\ Rev.\ C {\bf 90}, 015203 (2014).

\bibitem{nos_plb} V.~P.~Goncalves, B.~D.~Moreira and F.~S.~Navarra, Phys.\ Lett.\ B {\bf 742}, 172 (2015). 
 
\bibitem{armeza}  N.~Armesto and A.~H.~Rezaeian,
                  Phys.\ Rev.\ D {\bf 90}, 054003 (2014).
                      
\bibitem{Navelet}
  H.~Navelet and S.~Wallon,
  Nucl.\ Phys.\  B {\bf 522}, 237 (1998).


\bibitem{salam}
  A.~H.~Mueller and G.~P.~Salam,
  Nucl.\ Phys.\  B {\bf 475}, 293 (1996).


\bibitem{iancu_mueller}
  E.~Iancu and A.~H.~Mueller,
  Nucl.\ Phys.\  A {\bf 730}, 460 (2004); G.~P.~Salam,
  Nucl.\ Phys.\  B {\bf 461}, 512 (1996).
       
       
       
\bibitem{bkrunning}
  J.~L.~Albacete, N.~Armesto, J.~G.~Milhano and C.~A.~Salgado,
  Phys. Rev. D {\bf 80}, 034031 (2009). 


\bibitem{iim} E. Iancu, K. Itakura, S. Munier,  Phys. Lett. B {\bf 590}, 199  (2004).


\bibitem{soyez} G. Soyez,   Phys.\ Lett.\  B {\bf 655}, 32 (2007). 
       




\end{thebibliography}
\end{document}